\documentclass[prl,twocolumn,preprintnumbers,amsmath,amssymb]{revtex4}

\usepackage{graphicx}

\usepackage{amsmath,latexsym,epsf}

\def\be{\begin{equation}}
\def\ee{\end{equation}}

\def\bea{\begin{eqnarray}}
\def\eea{\end{eqnarray}}

\def\e{\epsilon}

\def\a{\alpha}

\def\k{\kappa}
\def\T{\Theta}
\def\t{\theta}
\def\p{\phi}

\begin{document}

\title{Correlation functions of one-dimensional anyonic fluids}
\author{Pasquale Calabrese and Mihail Mintchev}
\affiliation{Dipartimento di Fisica dell'Universit\`a di Pisa and INFN, 
Pisa, Italy}

\date{\today}

\begin{abstract}

A universal description of correlation functions of one-dimensional anyonic 
gapless systems in the low-momentum regime is presented.
We point out a number of interesting features, including universal
oscillating terms with frequency proportional to the statistical parameter 
and beating effects close to the fermion points.
The results are applied to the one-dimensional anyonic Lieb-Liniger model and 
checked against the exact results in the impenetrable limit.

\end{abstract}

\maketitle

One of the most striking results of modern physics is that in three spatial
dimensions particles are either bosons or fermions, i.e. the wave-function of
identical objects is either symmetrical or antisymmetrical under the
permutations of space coordinates.
Conversely, in lower dimensions generalized anyonic statistics, with  
intermediate properties between bosons and fermions are possible \cite{anyons}.
Fractional braiding statistics in two dimensions are considered a key
ingredient to describe fractional quantum Hall effect, motivating the large
scientific effort to understand two dimensional generalized statistics.
In one dimension (1D) collision is the only way to exchange particles and
consequently statistics and interaction are inextricably related.
This is the main reason why in 1D we can use bosonization and
fermionization (Jordan-Wigner) to attack complicated problems that in higher
dimensions are intractable. Excitations in 1D fluids can then
experience generalized statistics (not necessarily anyonic \cite{ges}).
However only recently a systematic study of interacting 1D anyonic systems 
started for different physical motivations. First transport in
quantum Hall fluids is localized on the edges and it is due to 1D
(chiral) anyons (see e.g. \cite{Fradkin}). Second recent experiments in
ultra-cold atoms succeeded in confining essentially 1D interacting systems
\cite{1datoms} and there is a proposal in this field to observe fractional
statistics \cite{paredes}.
However a general treatment of correlation functions of interacting anyonic
systems was still missing. In this letter we fill this gap giving a universal 
description of low-energy correlation functions in 1D anyonic fluids.

Anyonic statistics are described in terms of fields that at different points
($x_1\neq x_2$) satisfy
\bea
\Psi_A^\dag(x_1)\Psi_A^\dag(x_2)&=&e^{i\k \pi\e(x_1-x_2)} 
\Psi_A^\dag(x_2)\Psi_A^\dag(x_1)\label{ancomm}\,,\\
\Psi_A(x_1)\Psi_A^\dag(x_2)&=&e^{-i\k \pi\e(x_1-x_2)} 
\Psi_A^\dag(x_2)\Psi_A(x_1)\,,
\nonumber
\eea
where $\e(z)=-\e(-z)=1$ for $z>0$ and $\e(0)=0$. $\k$ is called statistical
parameter and equals $0$ for bosons and $1$ for fermions.
Other values of $\k$ give rise to general anyonic statistics ``interpolating''
between the two familiar ones. 

In 1D there is a natural way of defining an anyonic field in terms 
of a bosonic one \cite{Kundu}: 
$\Psi_A^\dag(x)=\Psi_B^\dag(x) e^{i\pi\k \int_{-\infty}^x dx' \rho(x')}$, where
$\rho(x)  \equiv \Psi_A^\dag(x)\Psi_A(x)=\Psi_B^\dag(x)\Psi_B(x)$, i.e.
anyonic and bosonic density are the same (there is also a construction in
terms of fermionic field \cite{Girardeau} that will not be exploited in this
letter). It is straightforward to show that if $\Psi_B$ satisfies usual bosonic
commutation relations, then $\Psi_A$ follows Eqs. (\ref{ancomm}).
In this letter we establish the behavior of correlation
functions of anyonic {\it gapless} systems (quantum fluids) in the low energy 
regime following 
the original idea of Haldane \cite{Haldane} for standard statistics (here 
we closely follow the notations of the review \cite{Cazalilla}). 
The starting point is to split $\rho(x)=\rho_<(x)+\rho_>(x)$, where 
$\rho_<$ refers to the slow parts (i.e. it consists of the fluctuations over
distances $\gg \rho_0=\langle \rho(x)\rangle$) and $\rho_>$ stands for the
remaining ``fast'' modes. On a physical basis it is natural to replace in the
definition of the anyonic field $\rho(x)$ with $\rho_<(x)$, since fast
fluctuations cancel in the integral.  
Then one can define the operator $\T(x)$ via $\partial_x\T(x)=\pi \rho_<(x)$
and the previously introduced anyonic field becomes
\be
\Psi_A^\dag(x)=\Psi_B^\dag(x) e^{i \k \T(x)}\,,
\label{anbosmap}
\ee
while the density is $\rho(x)=\rho_<(x)\sum_m e^{2mi\T (x)}$ 
\cite{Haldane, Cazalilla}.
In the bosonization (or equivalently conformal filed theory) language it is 
useful to introduce the operators $\p$
and $\t$ via $\t(x)=\T(x)-\pi \rho_0 x$ and 
$\Psi_B^\dag(x)= [\rho(x)]^{1/2} e^{-i\p (x)}$ 
which satisfy the commutation relation $[\p(x),\t(x')]=i \pi \e(x-x')/2$.
An alternative (constructive) approach would have been to start from these 
definitions of $\t$ and $\p$ and show that the field in Eq. (\ref{anbosmap})
satisfy anyonic statistics Eqs. (\ref{ancomm}), however we think that this way
of introducing the anyonic field is more pedagogical.

All correlation functions in the low momentum regime
can be obtained in terms of the so called vertex operators
\be
A_{m,n}(x)=e^{im\t(x)}e^{in\p(x)}\,.
\ee
For example, in terms of these operators it is straightforward to show that
when $|x|\gg \rho_0^{-1}$ the one-body density matrix (i.e. the Fourier
transform of the momentum distribution function) admits the expansion
\begin{multline}
g_1(x)=\langle\Psi_A^\dag(x)\Psi_A(0)\rangle\propto\\
\rho_0 \sum_{m=-\infty}^{\infty} e^{i(2m+\k)\pi\rho_0 x}
\langle A_{2m+\k,-1}(x)A_{-2m-\k,+1}(0)\rangle\,,
\end{multline}
that holds for all cyclic boundary conditions (the multiplicative factor is
determined only in terms of the high energy structure and it is not universal).
However to proceed we need to fix the boundary conditions (BC).
For simplicity, we now fix the bosonic field $\Psi_B$ to
satisfy periodic BC, obtaining (for a proof see e.g. \cite{Haldane, Cazalilla})
\be
\langle A_{2m+\k,-1}(x)A_{-2m-\k,+1}(0)\rangle\sim
\frac{e^{-i (2m+\k)\pi\e(x)/2}}{(\rho_0 d(x))^{\frac{(2m+\k)^2}2 K+\frac1{2K}}}\,,
\ee
that lead to the main result of this letter
\be
g_1(x)=
\rho_0 \sum_{m=-\infty}^{\infty} b_m
\frac{e^{i(2m+\k)\pi\rho_0 x}e^{-i (2m+\k)\pi\e(x)/2}  }{
(\rho_0 d(x))^{\frac{(2m+\k)^2}2 K+\frac1{2K}}}\,.
\label{main}
\ee
Here $d(x)=L\sin (\pi x/L)$, $b_m$ are unknown non-universal 
amplitudes that can be eventually fixed only through the exact/numerical 
computation of correlation functions in the specific model (as e.g. done for
the XXZ spin chain \cite{lt} and for the Bose gas \cite{ccs} using completely
different approaches). 
The exponent $K$ is instead universal and can be written 
as $K=\sqrt{v_J/v_N}$. $v_N$ and $v_J$ are phenomenological parameters (with
the dimensions of velocity) introduced by Haldane to fully characterize the
low-energy behavior of a quantum fluid \cite{Haldane}.
They correspond to density and phase stiffness respectively. 
In practice they must be extracted from the solution of the model (when 
available), from numerics or, ultimately, from experimental data.
In terms of these parameters, the velocity of the sound (i.e. the real speed
of excitations in the interacting system) is $v_s=\sqrt{v_J v_N}$.
For systems with Galilean invariance it holds $v_J=v_F$, where $v_F$ is the
Fermi velocity of a gas of spinless fermions with density $\rho_0$, and
consequently $K=v_F/v_s$.

Let us briefly discuss the role played by the BC.
Fixing periodic ones for the field $\Psi_B(x)$ (and also for $\t(x)$)
implies that $\Psi_A$ satisfy twisted BC 
$[\Psi_A^\dag(x+L)=e^{i\k\pi\rho_0 L} \Psi_A^\dag(x)]$. 
Requiring periodic BC for $\Psi_A$ leads to properly chosen twisted BC for
$\Psi_B$. Lack of space prevents us to show all these possibilities, but 
we note that when $\k \rho_0 L$ equals an even integer the two conditions are
equivalent. 
The appropriate BC for a given physical situation can be quite specific and 
depend on electrical and magnetic properties of anyons (see e.g. the example in
Ref. \cite{AB}). 
It is also easy to generalize the bosonic results of
Ref. \cite{Cazalilla} with open BC and at finite temperature to the present 
case. Asymptotic different times correlations can be obtained as well.
Note that, being the anyonic density equal to the bosonic one, the
density-density correlation function $g_2(x)=\langle\rho(x)\rho(0)\rangle$
coincides with Eq. (61) in Ref. \cite{Cazalilla} (obviously with $K$ fixed by
the anyonic dynamic).
All these applications of the present method will appear elsewhere
\cite{prep}.

Let us now discuss our finding in the thermodynamic limit $L\to\infty$. 
The sum Eq. (\ref{main}) is dominated by the smallest power law in the 
denominator. For $-1< \k<1$ this is always at $m=0$, and for large enough
systems we can neglect all the terms with $m\neq0$.
This results in a universal oscillating term $e^{i\k \pi\rho_0x}$ whose
frequency is a direct measure of the statistical parameter. 
Conversely for fermions with $\k=1$ the terms at $m=0$ and $m=-1$ display 
the same power law (and also, by symmetry, $b_0=b_{-1}$). 
Consequently, for $\k$ close to $1$ we expect beating effects
due to two oscillations with very close frequency and practically the same 
amplitude. 
The relevance of this effect will be explicitely showed later on.

Before considering specific applications, we make a last comment. The
definitions of anyonic fields (\ref{ancomm}) have a periodicity in $\k$ of 
period $2$.
However this is reflected in a periodicity of the system only if the
dynamics (dictated by some Hamiltonian) has the same period. In our approach
this is encoded in the exponent $K$ giving the real $\k$-periodicity
(e.g. in the Lieb-Liniger model discussed below the periodicity is $4$
\cite{Batchelor1}).

The previous derivation is completely general and holds for any interacting
1D anyonic fluid.
To show the predictive power of this analysis we now consider two specific
examples: the free gas and the Lieb-Liniger model.

{\it The free anyonic gas}.
The momentum distribution function of a (more general) gas of free anyons has 
been studied in the grand-canonical ensamble in Ref. \cite{Liguori}.
Considering only the case defined by Eqs. (\ref{ancomm}), fixing 
the chemical potential in terms of the mean-density, and performing a Wick
rotation we have
\be
g_1(x)\propto \frac{e^{i\k\pi(\rho_0x -\e(x)/2)}}{[L\sin (\pi x/L)]^\k}\,,
\ee
which is consistent with Eq. (\ref{main}) with $b_{m\neq0}=0$ and $K=\k^{-1}$
as known from the dispersion relation \cite{Liguori}.

{\it The Lieb-Liniger anyonic gas}.
An exactly solvable interacting model is the anyonic Lieb-Liniger gas, 
defined by the Hamiltonian:
\begin{multline}
H=\frac{\hbar^2}{2M}\int_0^L dx \partial_x\Psi_A^\dag(x) \partial_x\Psi_A(x)+\\
+c \int_0^L dx \Psi_A^\dag(x) \Psi_A^\dag(x) \Psi_A(x) \Psi_A(x)\,, 
\end{multline}
which describe $N$ anyons of mass $M$ on a ring of length $L$ interacting 
through a local pairwise interaction of strength $c$ (in what follow, we fix
$2M=\hbar=1$). The mean density is $\rho_0=N/L$.
For $\k=0$ the model reduces to the bosonic Lieb-Liniger \cite{LL}, while for 
$\k=1$ just consists of free fermions.
For $c=\infty$ it is equivalent to a gas of impenetrable anyons called
Tonks-Girardeau gas \cite{Girardeau}.
As in the bosonic case, the physics just depends on the dimensionless parameter
$\gamma=c/\rho_0$ and not on the two parameters separately.

The model is solvable by means of Bethe Ansatz \cite{Kundu, Batchelor1,Batchelor2}.
The Fock space is spanned by a set of Bethe wave functions, each fully
determined by a set of rapidities $k_j$ $j=1,\dots N$, solutions of 
the logarithmic Bethe equations (assuming anyonic periodic BC) 
\cite{Batchelor1}
\be
k_j L+\k\pi (N-1)+2\sum_{l=1}^N\arctan \frac{k_j-k_l}{c'}=2\pi I_j\,, 
\ee
in which $I_j$ are a set of integers (half-integers) quantum
numbers for $N$ odd (even), and $c'=c/\cos(\pi\k/2)$ 
is an effective 1D coupling of a bosonic equivalent model \cite{Kundu,foot1}.
If periodic BC are assumed for a corresponding bosonic field, the Bethe 
equations do not present the term $\k\pi (N-1)$ \cite{Kundu}.
Thus, a part from an eventual shift in the momentum, 
the Bethe equations are the same as the bosonic Lieb-Liniger at coupling $c'$: 
1D anyons with periodic BC are equivalent to bosons with twisted 
BC and viceversa, consistently with the previous general analysis.

For periodic BC, following \cite{Batchelor1}, we observe that any integer 
value of $\k (N-1)/2$ can be absorbed in the quantum number $I_j$ and only 
changes the correspondence between states and quantum numbers: 
$I_j'\equiv I_j-[\k (N-1)/2]$, where $[\cdot]$ stands for the integer part. 
For this reason it is worthy to define $\nu=\{\k (N-1)/2\}$, 
where $\{\cdot\}$ stands for the non-integer part $\{y\}=y-[y]$. 
It is then trivial to calculate the speed of the sound \cite{Batchelor1}
$v_s(\gamma,\k)= v_s(\gamma',0)+4\a\pi \nu/L$, where $\a=1(0)$ for periodic
(twisted) BC. Consequently 
$K(\gamma,\k)=v_F/v_s= 2\pi \rho_0/v_s(\gamma',0)+O(1/L)$.
In particular in the Tonks-Girardeau limit we always have 
$K(\gamma=\infty,\k)=1$.
For arbitrary $\gamma$, $v_s(\gamma,0)$ is not known analytically, 
but it is very easy to extract it numerically 
from the solution of the Bethe equations, see e.g. Ref. \cite{Cazalilla}. 

\begin{figure}[t]
\includegraphics[width=\columnwidth]{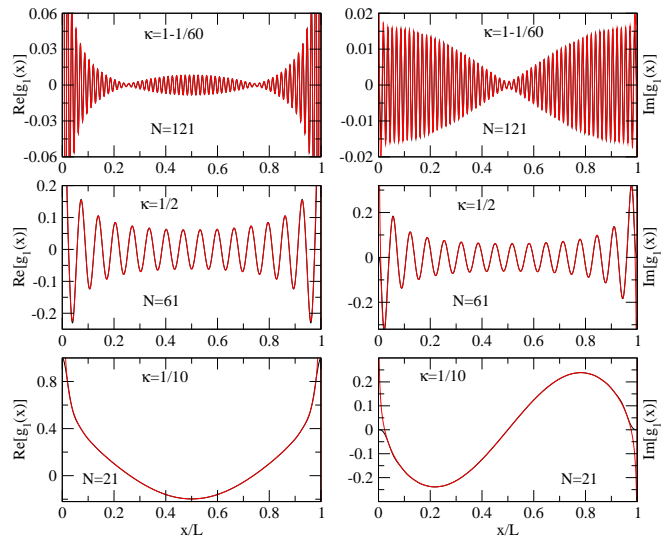} 
\caption{(Color online) 
Exact results for the anyonic Tonks-Girardeau gas (black lines)
of Ref. \cite{Santachiara} against the asymptotic large distance 
result (red lines) obtained here. }
\label{comparisonTG}
\end{figure}

Exploiting the anyon-fermion mapping Santachiara et al. \cite{Santachiara} 
gave an exact representation of the one-particle density matrix $g_1(x)$ in 
the Tonks-Girardeau regime ($c\to\infty$) in terms of the determinant of a 
Toeplitz matrix for cases when periodic BC are equivalent to twisted 
BC (i.e. $\kappa$ is an integer multiple of $2/(N-1)$). 
These exact results must be reproduced by our findings Eq. (\ref{main}) with
$K=1$. 
In Fig. \ref{comparisonTG} we compare the results of Ref. \cite{Santachiara}
with Eq. (\ref{main}) for the same values of $\k$ plotted in 
Ref. \cite{Santachiara}. 
For $\k=1/10$ ($N=L=21$) and $\k=1/2$ ($N=L=61$) we set $b_{m\neq0}=0$ 
and we fix $b_0$ as an overall normalization by imposing the correct 
value at $x=L/2$.
The value $\k=59/60$ ($N=L=121$) is peculiar because very close to the 
free-fermionic point $\k=1$, where two power-laws ($m=0,-1$) 
have the same exponents, $b_0=b_{-1}$ and all the other $b_m=0$.
For $\k=59/60$ and $N=121$ the two power laws are still indistinguishable,
thus to compare the exact results with Eq. (\ref{main}) we fix $b_0=b_{-1}$ 
and all the other $b_m=0$. 
The beating effect resulting from two oscillations with almost the
same amplitudes is evident as overall nodes (in the amplitude of the fast
oscillations) in the real part of $g_1(x)$ at 
$x/L=1/4$ and $3/4$, in contrast with the monotonic behavior for smaller $\k$
(see Fig. \ref{comparisonTG}).
In the three cases the agreement between exact and asymptotic
results is perfect.

Moreover in Ref. \cite{Santachiara} the asymptotic behavior for large $N$ 
of $g_1(x)$ was estimated using Fisher-Hartwig conjecture.
The results obtained in this way are equivalent to the zero mode ($m=0$) in
Eq. (\ref{main}) \cite{foot2}.
On the one hand our results give further evidence that the Fisher-Hartwig 
conjecture should be valid and is able to fix properly the constant $b_0$ 
that in our approach remains unknown.
On the other hand, Eq. (\ref{main}) shows that, even for very large $L$, close 
to the Fermi point the resonance effects cannot be caught by other asymptotic
approaches (compare the accuracy of the plot in Fig. \ref{comparisonTG} for
$\k=59/60$ with the analogous one in Ref. \cite{Santachiara}). 
However, hard-core anyons are only a specific case to which apply
Eq. (\ref{main}), on the same foot as the Lenard formula \cite{Lenard} for
impenetrable bosons is only the simplest case described by the hydrodynamic 
treatment of Haldane \cite{Haldane} for standard statistics.

{\it Discussions}. 
We provided a universal description of correlation functions 
of 1D anyonic fluids in the low-momentum regime.
Our main result is the zero-temperature one-particle density matrix 
given by Eq. (\ref{main}).
This formula unambiguously shows how to connect smoothly correlations of 
{\it interacting} bosons and fermions through the continuous parameter $\k$.

We specifically applied our method to the Lieb-Liniger anyonic gas, but it
would be interesting to understand the consequences of our findings in more
complicated fluids, as for example a generalized Tomanaga-Luttinger 
liquid \cite{prep}, fractional generalization of Heisenberg spin chains as,
e.g., the golden chain \cite{Feiguin}, fluids with internal degrees of freedom
(e.g. anyonic extensions of the model in \cite{Yang}) etc..

We find that the main universal signature of anyonic statistics in the 
correlation functions 
is the oscillating complex term that can lead to clear signals in
transport experiments \cite{Fradkin}.
The remaining part in Eq. (\ref{main}) is very similar to the bosonic
counterpart \cite{Haldane} with the generalized statistics leading to a
predictable ``renormalization'' of the power-law behavior for large $x$.
This is a manifestation of the 1D interplay between interaction and
statistics in the low momentum regime. It would be extremely interesting to
understand to which extent some analogous correspondence can be valid for all
momenta and in gapped systems.

{\it Acknowledgments}. We thank J.-S. Caux, N. Magnoli and R. Santachiara for 
very fruitful discussions. In particular we are indebted with R. Santachiara
and F. Stauffer for providing us numerical data in Ref. \cite{Santachiara}.


\begin{thebibliography}{99}
\bibitem{foot2}
An error in the computation of the exponent of $d(x)$ was corrected thanks to
the comparison with our findings, see v3 of Ref. \cite{Santachiara}.
\end{thebibliography}
\end{document}